\documentclass[%
 amsmath,amssymb, 
 aps,pra,twocolumn,superscriptaddress
]{revtex4-1}

\usepackage{graphicx}
\graphicspath{{Figures/}}

\newcommand{\id}{\mathbf{1}}
\usepackage{physics}
\usepackage{subfiles}
\usepackage{multirow}
\usepackage{listings}
\usepackage{color}
\usepackage[version=4]{mhchem}

\begin{document}

\title{Analysis of Superfast Encoding Performance for Electronic Structure Simulations}

\author{Riley W. Chien}
\affiliation{Department of Physics and Astronomy, Dartmouth College, Hanover, New Hampshire 03755}

\author{Sha Xue}
\affiliation{Department of Physics and Astronomy, Dartmouth College, Hanover, New Hampshire 03755}

\author{Tarini S. Hardikar}
\affiliation{Department Chemistry, University of California, Berkeley, California 94720}

\author{Kanav Setia}
\affiliation{Department of Physics and Astronomy, Dartmouth College, Hanover, New Hampshire 03755}

\author{James D. Whitfield} 
\affiliation{Department of Physics and Astronomy, Dartmouth College, Hanover, New Hampshire 03755}

\date{\today}
\begin{abstract}
In our recent work, we have examined various fermion to qubit mappings in the context of quantum simulation including the original Bravyi-Kitaev Superfast encoding (OSE) as well as a generalized version (GSE). We return to OSE and compare it against the Jordan-Wigner (JW) transform for quantum chemistry considering the number of qubits required, the Pauli weight of terms in the transformed Hamiltonians, and the $L_1$ norm of the Hamiltonian. We considered a test set of molecular systems known as the Atomization Energy 6 (AE6) as well as Hydrogen lattices. Our results showed that the resource efficiency of OSE is strongly affected by the spatial locality of the underlying single-particle basis. We find that OSE is outperformed by JW when the orbitals in the underlying single-particle basis are highly overlapping, which limits its applicability to near-term quantum chemistry simulations utilizing standard basis sets. In contrast, when orbitals are overlapping with only few others, as is the case of Hydrogen lattices with very tight orbitals, OSE fares comparatively better. Our results illustrate the importance of choosing the right combination of basis sets and fermion to qubit mapping to get the most out of a quantum device when simulating physical systems.
\end{abstract}

\maketitle

\twocolumngrid
\section{Introduction}

To utilize the power of quantum computation for electronic structure theory (see \cite{mcardle2018,GeorgescuNori}, and references therein), it is crucial to map fermionic modes to qubits efficiently. Efficient mapping allows the number of overall quantum gates applied to be reduced, which is especially important given the current limitations of quantum computing hardware due to decoherence and noisy gates. While there are multiple fermion to qubit encodings \cite{Bravyi2000,Seeley2012,Verstraete2005,Ball2005,Steudtner2017}, recent work has focused on the potential of the Bravyi-Kitaev Superfast encoding for quantum simulation \cite{Bravyi2000,Havlicek2017,Setia2018}. Since we have also developed a generalization of the Bravyi-Kitaev Superfast encoding, called the Generalized Superfast Encoding (GSE) \cite{Setia2018b}, we opt to refer to the original Superfast encoding as OSE.  
In some settings, OSE may offer advantages over other mappings such as the JW mapping or the Bravyi-Kitaev mapping (a separate mapping which can be thought of as compromise between directly storing occupation and parity). 
This potential is due to the fact that the relevant quantities (qubits, gates) scale as a function of the number of interactions between orbitals, rather than the number of orbitals themselves \cite{Havlicek2017}. However, the performance of OSE compared to other mappings in the context of electronic structure problems has yet to be explored.  Our work here testing the OSE against JW can be compared to Ref. \cite{tranter2018} where the Bravyi-Kitaev transform and the JW transformations were compared. Both the JW and the Bravyi-Kitaev use the same number of qubits as spin-orbitals whereas the OSE and the GSE, in general, utilize different qubit resources. Here, we present the results of implementing OSE, alongside JW, for various molecular systems, including Hydrogen lattices and chains, and molecules from a small test set \cite{Lynch2003}. 

In this study, we are aiming to characterize the performance of standard quantum simulation techniques based on both hybrid quantum-classical and time-propagation methods. Hence, we used two main metrics to determine the performance of an encoding on a given fermionic Hamiltonian. The first is simply the number of qubits required to simulate the system after performing the mapping. This is an important metric given that on near-term devices the number of qubits is severely limited. Second, we also considered the tensor weights of the transformed Hamiltonian. This is defined as the number of all non-identity tensor factors of a Pauli term. The total tensor weight is the sum of all weights in the transformed Hamiltonian. We also looked at other statistics such as the average weight and the maximum tensor weight among terms. As a metric of performance we believe the total tensor weight is reasonable because the tensor weight will affect the ultimate circuit depth of the quantum simulation algorithm. We note, however, that it is not exactly analogous given the fact that there will be terms which commute with each other and would be able to be performed in parallel. In addition, different techniques can be used to simulate the Hamiltonian evolution such as Trotterization, truncated Taylor expansions, and qubitization, see e.g. \cite{Berry2015, Low2016}. The choice of simulation technique will ultimately affect the circuit depth, however the tensor weight is unambiguous. Finally, motivated by recent developments we looked at two $1$-norms which have been shown to give a bound on the scaling of various Hamiltonian simulation algorithms \cite{campbell2018random}. In particular, we focused on the coefficient $L_1$ norm defined as:
\begin{equation}
\|A\|_c = \sum_j |c_j| \quad\textrm{ for } A=\sum_j c_j P_j .
\end{equation}
In the coefficient $L_1$ norm, coefficients are prefactors of a decomposition into Pauli group elements $P\in \{ \prod_j \sigma^\mathbf{m}_j : \mathbf{m}\in\{x,y,z,0\}\}$. It is also an upper bound to the $L_1$ induced norm $\|A\|_1 = {\displaystyle \max_{x \neq 0}} |Ax|_1/|x|_1 =\max_j \sum_i |a_{ij}|$. 

The calculations were performed using three different basis set transformations: the canonical Hartee-Fock molecular orbitals, the symmetrically orthogonalized atomic orbital basis, and the canonically orthogonalized atomic orbital basis which we discuss later in Section \ref{ssec:basis_rotations}. The calculations on the Hydrogen systems were run using a highly localized Gaussian basis set employing symmetric orthogonalization, while the AE6 molecules used the STO-3G basis set \cite{Pople1969, Pople1970}. In the Hydrogen systems, using a tight Gaussian basis, the performance of OSE was mixed and depended on dimension. In one dimension, the number of required qubits was reduced at the cost of a larger total tensor weight. In higher dimensions, the total number of qubits required is larger than JW but the scaling of the largest tensor term is improved - a feature that, at system sizes larger than we were able to numerically explore for this study, is expected to lead to a crossover in the total number of gates. 
The reason for the dependence on dimension is the interaction-based nature of OSE, which is explored in greater detail below as well as in full detail in \cite{Setia2018}.

Previous work \cite{Havlicek2017} has predicted that for the Hubbard model, the gate count for OSE will scale more favorably than JW. As we will discuss, our Hydrogen lattice results hint at agreement with this prediction, however for the small size Hydrogen lattices that we investigated here JW still performs better than OSE.
For the AE6 molecules using STO-3G, JW outperformed OSE in all respects. This suggests that broad Gaussian basis sets centered on atomic nuclei are not practicable when used with OSE and another mapping such as JW would be preferable for such basis sets. Using OSE with localized basis sets shows interesting potential for electronic structure and solid state calculations, and it is hoped that this mapping can be tested on present quantum computers \cite{Babbush2018,Kivlichan2018}. Our results illustrate the importance of choosing the right combination of basis sets and fermion to qubit mapping to get the most out of a quantum device when simulating physics and chemistry. 

The remainder of this paper is organized as follows. In Section \ref{sec:background}, we first provide an overview of fermion to qubit mappings including JW transformation and OSE, and the single particle basis set rotations. Then, we describe the molecules we studied and the computational method and metrics for comparison we used. In Section \ref{sec:analysis}, we discuss some analytic results including scaling in the case of complete graphs as well as expected results using atom-centered basis sets. In Section \ref{sec:numerics}, we present and discuss the numerical results for the Hydrogen systems and the AE6 molecules. A comparison between JW and OSE with different basis rotations is given. Finally, we summarize the conclusions based on our investigations. The notation conventions used in this paper are $N$ for the length of Hydrogen chains in atoms, $M$ the number of spin-orbitals, $Q$ denotes the number of qubits, and indices start from $0$.  We will be using atomic energy units where the mass of the electron, the electronic charge, Planck's constant and $(4\pi\epsilon_0)^{-1}$ are all set to unity.

\section{Background}\label{sec:background}

\subsection{Jordan-Wigner}
There exists a useful correspondence between fermionic modes and qubits, since the occupation number of a particular orbital can be associated with the two states of the qubits. Direct identification between occupation of orbitals and the two levels of qubits, was first defined in the context of 1-D lattice models \cite{Jordan1928} and then proposed as a scheme for simulating fermions \cite{Ortiz2001}. 

The raising and lowering operators for qubits, given as:
\begin{align}
\sigma^{\pm}=\frac{1}{2}(\sigma^{x}\mp i\sigma^{y})
\end{align}
do not in general satisfy the anti-commutation relations satisfied by fermions: 
\begin{align}
\{a_j,a_k\}&=a_ja_k+a_ka_j=0,
&\{a_j,a_k^{\dag}\}&=\delta_{jk}\label{ac_relations}
\end{align}
where, $a_i$ and $a_i^{\dagger}$ are the annihilation and creation operators acting upon fermionic mode $i$. In order to satisfy the necessary anti-commutation relations, a string of Pauli $\sigma^z$ operators are included. This then gives the necessary ingredients for the transformation which is defined as follows:
\begin{align}
a_j^{\dag}\equiv\mathbf{1}^{\otimes M-j-1}\otimes \sigma_{j}^{+}\otimes[\sigma^z]^{\otimes j}\\
a_j\equiv\mathbf{1}^{\otimes M-j-1}\otimes \sigma_{j}^{-}\otimes[\sigma^z]^{\otimes j}
\end{align}
Under this transformation, the fermionic Hamiltonian represented in second quantization can be mapped to a Hamiltonian acting on a qubit Hilbert space \cite{Seeley2012,Havlicek2017}.  Note that the string of Pauli $\sigma^z$ operators included to satisfy the anti-symmetry properties of the fermionic operators can in some cases act on a large number of qubits. Thus necessitating many quantum gates to simulate the system under consideration. There have been techniques devised to mitigate this problem including the introduction of auxiliary qubits \cite{Steudtner2017}. Described below, OSE also helps in reducing this overhead.

\subsection{OSE Mapping}
OSE is a mapping from fermionic operators to qubit operators which is in essence, based on the interactions between fermionic modes rather than the occupancy of the modes as in JW. An interaction graph is defined based on the interaction between modes in the Hamiltonian which is to be encoded. The vertices of the graph correspond to the modes. If there is a term in the fermionic Hamiltonian which couples two modes, an edge connecting the two corresponding vertices is included in the graph. Qubits are then identified with the edges of the graph and operators which act on the qubits are defined. In particular, there are two types of operators: edge operators which act on qubits in a small neighborhood of the edge in question and vertex operators which act on all edges incident to the vertex in question. The transformation from fermionic operators to qubit operators is then defined by the edge and vertex operators. 

First, the fermionic modes are expressed as Majorana modes
\begin{align}
c_{2j}&= a_{j}+a^{\dagger}_{j} \\
c_{2j+1}&= -i(a_{j}-a^{\dagger}_{j})
\end{align}
then edge and vertex operators are defined over the Majoranas
\begin{align}
	B_j&=-i c_{2j}c_{2j+1} \quad \text{for each vertex,}\\
	A_{jk}&=-i c_{2j}c_{2k} \quad \text{for each edge }(j,k) \in E. \label{eq:edge_op}
\end{align}
These edge and vertex operators satisfy the algebra:
\begin{align}\label{eq:algebra1}
    B_i &= B_i^{\dag}, & A_{ij} &= A_{ij}^{\dag}\\
    B_i^2 &= \id, & A_{ij}^2 &= \id \label{eq:algebra2} \\
    B_i B_j &= B_j B_i, & A_{ij}&= -A_{ji} \label{eq:algebra3}
\end{align}
\begin{align}\label{eq:algebra4}
    A_{ij} B_k &= (-1)^{\delta_{ik}+\delta_{jk}}B_k A_{ij}\\
    A_{ij} A_{kl} &= (-1)^{\delta_{ik}+\delta_{il}\delta_{jk}+\delta_{jl}}A_{kl}A_{ij}\label{eq:algebra5}
\end{align}
\begin{equation}\label{eq:loops}
    i^p A_{j_0 j_1}A_{j_1 j_2}\dots A_{j_{p-2} j_{p-1}}A_{j_{p-1} j_0}=\id
\end{equation}
\begin{center}
    for any closed loop of $p$ edges on the graph.
\end{center}

\begin{table*}
    \centering
    \caption{Second quantized molecular Hamiltonian operators in terms of edge operators in the original Superfast encoding (OSE) and the expressions for edge operators in terms of qubit Pauli operators (see \cite{Setia2018}). In the edge operator expression, $n(j)$ denotes the set of indices corresponding to neighbors of vertex $j$ and $\epsilon_{pq} = +1 \text{ for } p < q$ and $ \epsilon_{pq}=-1 \text{ for } p>q $.}
    \begin{tabular}{c c c}
        \hline
        \textbf{Operator type}  &\phantom{space} \textbf{Second quantized form} \phantom{space} & \textbf{OSE form}\\
        \hline
        Pair creation operator & $a_i^{\dagger}a_j^{\dagger} + a_i a_j $ & $-i(A_{ij}B_j - B_i A_{ij} )/2$ \\[2ex]
        Number operator & $h_{ii}a_i^{\dagger}a_i$  & $h_{ii}(1-B_{i})/2$\\[2ex]
        Coulomb/exchange operator & $h_{ijji}a_i^{\dagger}a_j^{\dagger}a_j a_i $ & $ h_{ijji}(1-B_i)(1-B_j)/4$\\[2ex]
        Excitation operator & $h_{ij}(a_i^{\dagger} a_j +a_j^{\dagger} a_i )$ & $-i h_{ij}(A_{ij}B_j + B_i A_{ij})/2 $\\[2ex]
        Number-excitation operator & $h_{ijjk}(a_i^{\dagger}a_j^{\dagger} a_j a_k + a_k^{\dagger}a_j^{\dagger} a_j a_i) $ & $-i h_{ijjk}(A_{ik}B_k + B_i A_{ik})(1-B_j)/4 $\\[2ex]
        Double-excitation operator & $h_{ijkl}(a_i^{\dagger}a_j^{\dagger} a_k a_l + a_l^{\dagger}a_k^{\dagger} a_j a_i)$ & $h_{ijkl} A_{ij} A_{kl}(-1-B_i B_j + B_i B_k+ B_i B_l + B_j B_k$\\
        \qq & \qq & $+ B_j B_i - B_k B_l - B_i B_j B_k B_l)/8$ \\ [2ex]
        \hline
        \textbf{Operator type} & \textbf{OSE operator} & \textbf{Pauli representation}\\[1ex]
        \hline
        Vertex operator & $B_i$ &$\displaystyle \prod_{j: (ij)\in E}\sigma^{z}_{ij}$ \\[4ex]
        Edge operator & $A_{pq}$ & $\displaystyle\epsilon_{pq} \sigma^{x}_{pq}\prod_{l<q}^{n(p)} \sigma^{z}_{lp} \prod_{s<p}^{n(q)} \sigma^{z}_{sq}$\\[2ex]
        \hline
    \end{tabular}

    \label{table:OSEterms}
\end{table*}
The qubit representation of the edge and vertex operators is given in TABLE \ref{table:OSEterms}. These qubit operators satisfy all of the conditions in (\ref{eq:algebra1}-\ref{eq:algebra5}) however they do not in general satisfy the loop condition (\ref{eq:loops}). The loop condition is however satisfied in a subspace of the total Hilbert space so we use this condition to specify a restricted subspace which we call the code space which is stabilized by the loop operators. Restricting to the code space completes the transformation. Since the encoded fermionic operators commute with the loop operator, once the system is initialized into the code space, time-evolution according to the encoded Hamiltonian will only occur within the code space. As is suggested by the presence of stabilizers, OSE carries error correcting properties, namely, that if the degree of the interaction graph is at least six, then it is possible to correct single-qubit errors \cite{Setia2018}. The molecular Hamiltonian operators are presented along with their expression in terms of edge operators in TABLE \ref{table:OSEterms}.

\subsection{Rotations of the single-particle basis set}\label{ssec:basis_rotations}

In this work, we fix the single particle basis set e.g. STO-3G \cite{Hehre69} as well as customized basis sets but consider three different rotations of the given basis set.  The single particle functions are in general non-orthogonal with an overlap matrix possessing an eigendecomposition $S=UsU^\dag$ and $S_{ij}=\bra{i}\ket{j}$. We have considered three procedures to orthogonalize them: the canonical (AOC), the symmetrical (AOS), and the Hartree-Fock molecular orbital bases. 
While we considered three, there are an infinite number of possible bases to chose parameterized by $X=Us^{-1/2}W$ for any unitary matrix $W$.  If $W=\id$ then the basis is called the canonical orthogonalized basis while if $W=U^\dag$ the basis is symmetrically orthogonalized. Beginning with the eigendecomposition of $S$, the corresponding orthogonalizing matrices are $X^S=S^{-1/2} = Us^{-1/2} U^\dagger$ and $X^C=U s^{-1/2}$ for AOS and AOC respectively.

In our calculations on Hydrogen chains and lattices we give particular focus to the symmetrical orthogonalization due to its important feature that the resulting othogonalized functions are those that minimize the $L_2$ distance between the initial and final sets of basis functions \cite{Carlson1957}. This feature is especially attractive as we have chosen basis sets which seek to minimize the overlaps and preserving locality is essential.

However, note that when the basis set, here $\{\phi_i\}$, has linear dependence or near linear dependence, eigenvalues will approach zero, and might be small enough to lead to machine precision errors in calculating $s^{-1/2}$. Canonical Orthogonalization is used to fix this problem. Similar to symmetric orthogonalization, if the basis set is nearly linear dependent, then eigenvalues of $S$ can approach zero, causing $s^{-1/2}$ to not be ill-defined. In this case, the eigenvalues which cause machine precision errors can be removed and a truncated matrix $\tilde{X}$ can be constructed. In $\tilde{X}$ only  eigenvectors corresponding to the non-trivial eigenvalues are kept \cite{Szabo1996}.

\begin{figure}
    \centering
    \includegraphics[width=7cm]{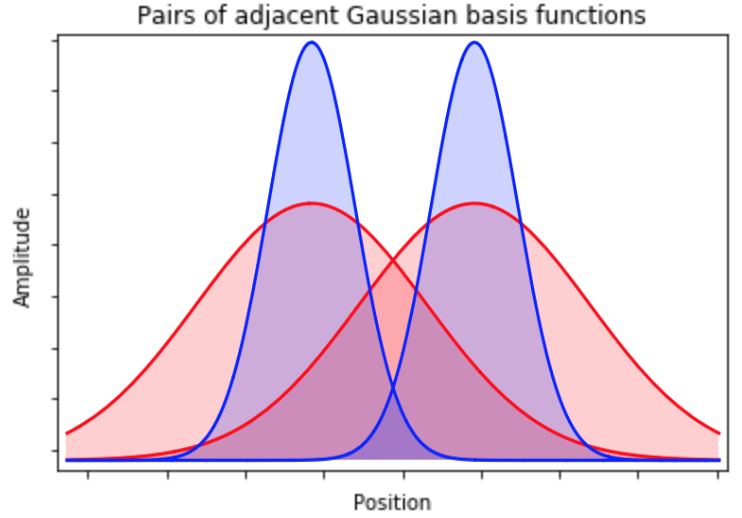}
    \caption{Overlap between pairs of Gaussian orbitals of different widths centered on neighboring sites $1$ Angstrom apart.}
    \label{fig:Gaussians}
\end{figure}

\subsection{Molecules studied}
Molecular test sets containing experimentally validated results and values are often used to test new developments in electronic structure theory. This is especially important in the context of density functional theory \cite{Goerigk17,Ramakrishnan14} where there is not a systematic way to improve the results of a given function.  Other newly developed methods are also tested using test sets.  

Thus far, quantum computing for quantum chemistry has relied heavily on the Hydrogen molecule as a prototypical example.  Here we move beyond molecular Hydrogen to examine a small test set of six molecules.  The set we look at, named AE6, was first introduced in \cite{Lynch2003} as a selected representative of a much larger database of electronic structure test instances.  The AE6 testset includes: Silane (\ce{SiH_4}), Sulfur (\ce{S_2}), Propyne (\ce{C_3H4}), Silicon Monoxide (\ce{SiO}), Glyoxal (\ce{C_2H_2O_2}), and Cyclobutane (\ce{C4H8}) with all input geometries extracted from \cite{Lynch2003}.

For the purposes of theoretical simplicity, we also explored model metallic Hydrogen chains and lattices with nuclei spaced one Angstrom apart on a rectilinear grid. 
Similar studies of Hydrogen chains were used as a simple model and as an algorithmic benchmark for state-of-the-art numerical methods \cite{Motta2017}. 
With these two sets, we are aiming to benchmark the use of OSE in theoretically simply settings and in standard deployment areas for electronic structure methods in molecular physics.

\subsection{Software and Setup}
This work was done using Psi4, an open source software for quantum chemistry \cite{Psi42017}, and OpenFermion \cite{McClean2017}, an open source software for quantum fermionic mappings. A plugin, developed by the OpenFermion developers, ``OpenFermion-Psi4" was used to interface the two.

\section{Analytic Results}\label{sec:analysis}
\subsection{Complete Graph Upper Bounds}
Here, we look at the largest qubit operators generated by OSE as a function of the number of spin-orbitals $M$ when the graph (pair of graphs one for each spin sector) is complete. This will be the case in which OSE is likely not the ideal choice of mapping but it is instructive to see the effect that, for example, many tightly packed basis functions has on the mapping.
Suppose that corresponding to each spin sector, we have a complete graph of fermionic modes, $K_{M/2}$.
We then have $2\tbinom{M/2}{2}$ edges present in the graph which is also the number of qubits required. 
Vertex operators $B_i$ will consist of a product of $\sigma^z$ operators one for each edge incident to vertex $i$ giving a weight $M/2-1$.
Edge operators $A_{ij}$ are of variable length depending on the indices $i,j$. They consist of a product of one $\sigma^x$ and two strings of $\sigma^z$ operators each of average length $M/4$.
Fermionic double-excitation operators $a_i^{\dagger}a_j^{\dagger}a_k a_l$ will on average be mapped to the Pauli terms with the largest tensor weight and they are also the most numerous provided that none of the integrals fall below the specified cutoff. Thus, we will consider those operators in determining the polynomial scaling. There will be $O(M^4)$ such operators present in the second quantized Hamiltonian. Double excitation operators are mapped to a sum of products of edge and vertex operators, having the form given in TABLE \ref{table:OSEterms}.
The largest weight term in the sum is typically $A_{ij}A_{kl}B_i B_j B_k B_l$. Due to cancellations in the $\sigma^z$ strings between $A_{ij}$ and the vertex operators $B_i$ and $B_j$ there will on average be $M/2$ such $\sigma^z$ tensor factors being contributed from the $i,j$ indices as well as another $M/2$ from the $k,l$ indices so the weight of each double excitation operator scales linearly as $O(M)$. The worst-case scaling on the total tensor weight for an OSE fermionic Hamiltonian of $M$ modes is therefore $O(M^5).$ This worst-case scaling is in line with previous results for estimating the gate cost scaling of simulating quantum chemistry with JW \cite{Wecker2014}. 

\subsection{Atom-centered Basis Sets Lower Bounds}
For a given atom-centered basis set, there exists a lower bound for the number of qubits required. As the basis functions become more localized, interaction terms between orbitals disappear, therefore, the number of qubits required decreases. However, even if there are no interactions between modes centered on different atoms, the interactions between orbitals centered on each atom persist. This leads to a lower bound on the number of edges needed within molecules. This lower bound for the atom-centered basis sets can be expressed as:
\begin{align}
   &Q_L = 2 \sum_{a=atom}\tbinom{M_a}{2} 
\end{align}
 with $M_a$ as the orbitals (spin up or spin down) on atom $a$. Take SiO as an example. There are 14 spatial orbitals (28 spin orbitals) which include 9 orbitals from Si and 5 orbitals contributed by O. With very tight orbitals, there is no differential overlaps, therefore, no edges between the orbitals for Si and orbitals for O. The total edges for SiO molecule would be $2 \tbinom{9}{2}$ + $2 \tbinom{5}{2}$ = 92. The lower and upper bounds for AE6 molecules are summarized in TABLE \ref{table:bounds}. We also include qubits for JW as comparison. From TABLE \ref{table:bounds}, it is clear that the qubits required by JW are much fewer than OSE for the AE6 molecules. A numerical investigation of the relationship between qubit requirements and localized orbitals follows in the next section.

\begin{table}
\centering
\caption{OSE upper and lower bounds on the number of qubit required for AE6 molecules denoted by $Q_U$ and $Q_L$ respectively. The JW qubit number is included for comparison.}

\begin{tabular}{c|ccc}
 \hline
   Molecules   &  $Q_U$   &   $Q_L$   &   $Q_{JW}$  \\ 
 \hline
 Silane & 156 & 90 & 26 \\ 
 SiO & 182 & 92 & 28 \\
 Sulfur & 306 & 180  & 36 \\
 Propyne & 342 & 90  & 38 \\
 Glyoxal & 462 & 120  & 44 \\ 
Cyclobutane & 756 & 120  & 56 \\
 \hline

\end{tabular}
\label{table:bounds}
\end{table}

\subsection{Empty graphs and odd-parity state preparation}\label{sec:empty}

It is possible through the choice of very narrow basis functions or an exceptionally high cutoff to have a Hamiltonian which takes the diagonal form of $H_d=\sum_j h_j n_j +\sum_{ij} W_{ij}n_in_j$. Observing the transformation of these terms as found in TABLE \ref{table:OSEterms}, we note that no $A_{pq}$ terms are required by the interactions.  Thus the Hamiltonian requires no edges to simulate the interactions.  Although in the case of zero qubits, one should think of the only possible state is that of the vacuum fermionic state.  To prepare additional states i.e. $a_m^\dag a_n^\dag\ket{\Omega}$, then edge $(m,n)$ must be included.  Once additional edges are present, the Hamiltonian $H_d$ can then determine the energy of the distinct states.  


A second challenge faced by this mapping is the restriction to even and odd particle number sector. 
Note odd products of fermionic operators can not be constructed from edge operators and the vacuum state. If a particle is desired in mode $k$, an ancillary fermionic mode, $s$, would have to be introduced and an edge placed between the vertex corresponding to the desired mode and the ancillary one. The edge is then acted upon by the pair creation operator:
\begin{align}
a_k^{\dagger}a_{s}^{\dagger}+ a_{s} a_k = \frac{-i}{2}(A_{k,s}B_{s} - B_k A_{k,s}) .
\end{align}
This is the way in which states with odd parity can be constructed. The ancillary mode then takes no part in dynamics generated by the Hamiltonian.

\section{Numerical Results}\label{sec:numerics}

\subsection{Hydrogen Chains and Lattices}

Due to the interaction-based construction of OSE, the setting in which it offers possibly the greatest potential is the simulation of systems in which the fermionic modes are coupled to few others especially those involving nearest neighbor couplings. A classic well-studied fermionic system of nearest-neighbor interactions on a lattice exhibiting rich physics is the Hubbard model involving hopping terms and Coulomb terms and its simulation on a quantum computer has previously been studied \cite{weckerhubbard}. The Hubbard model can be interpreted as a chain or lattice of Hydrogen atoms under an approximation neglecting differential overlaps, setting $h_{ijkl} = \gamma_{ij}\delta_{il}\delta_{jk}$ \cite{schatz2002quantum}. 

In order to investigate the performance of the encoding in systems with local interactions, we have considered configurations of Hydrogen atoms in 1-, 2-, and 3-dimensions. We have made the simplification to limit to a single Gaussian basis function. Symmetric orthogonalization is also used to preserve locality. Note that our intention here was not to examine the performance of 
OSE in realistic molecular systems as we have done with the AE6 molecular test set.  Rather, we are attempting to illustrate in explicit detail the dependence on the degree and dimension in the performance of the mapping as applied to lattice systems. To that end, we have varied the Gaussian basis functions from broad to narrow in their spatial extent.  When the orbitals are delocalized many modes are interacting resulting in an interaction graph of relatively high degree.  When the Gaussians are narrow and sharply-peaked interactions other than nearest neighbor are negligible at the cutoff of $10^{-7}$. The transition from tight to broad Gaussian orbitals appears clearly in the data as a large increase in both number of qubits as well as total operator tensor weight.
\begin{figure*}
    \centering
    \includegraphics[width=17cm]{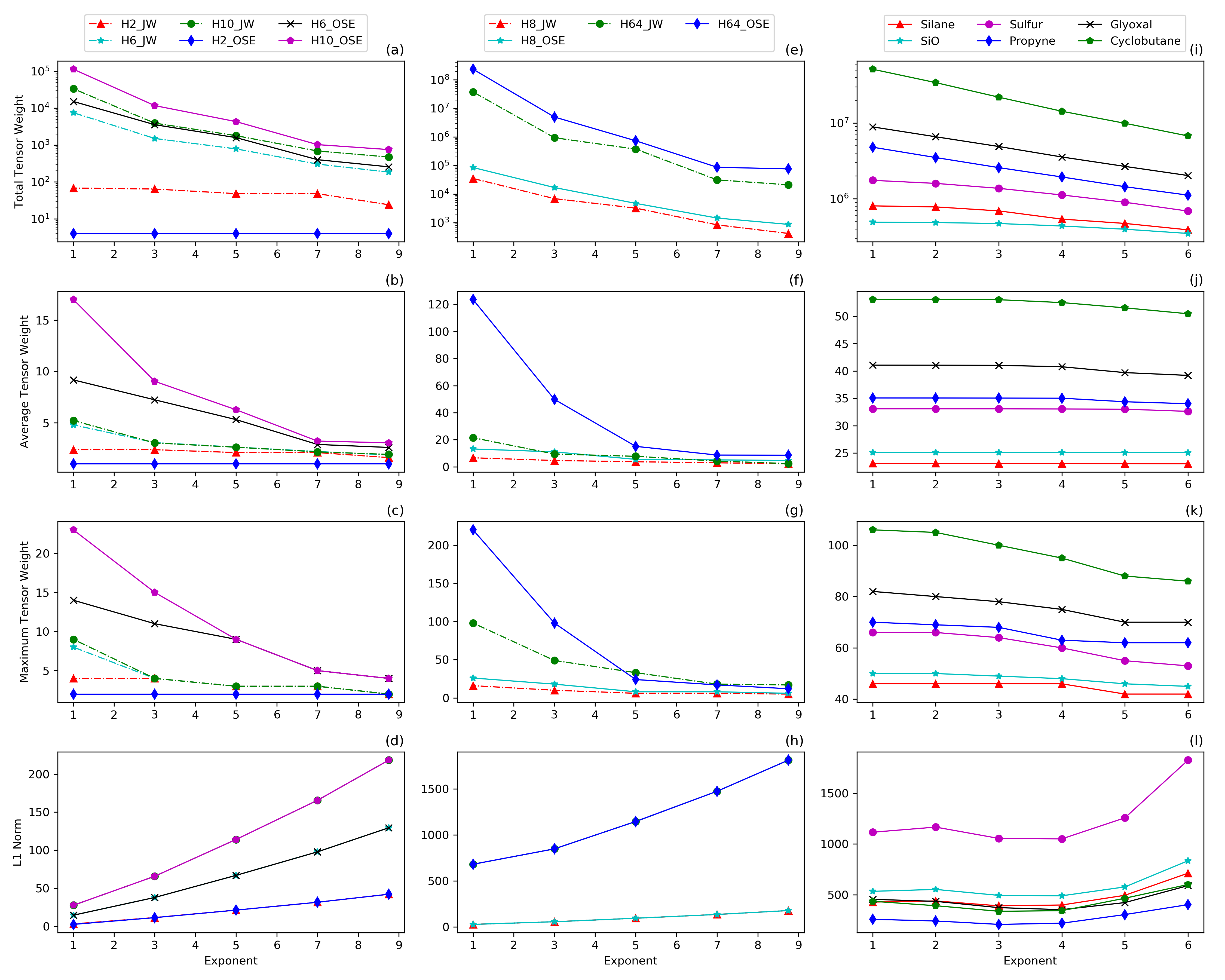}
    \caption{Numerical results (Total Tensor Weight, Average Tensor Weight, Maximum Tensor Weight and $L_1$ Norm in JW and OSE transformed qubit Hamiltonians with different customized basis sets) for Hydrogen chains and AE6 molecules. First column (a)-(d): 1-D Hydrogen chain. Second column (e)-(h): 3-D cubic lattice of Hydrogen atoms. Third column (i)-(l): AE6 molecules (OSE only).}
    \label{fig:numericalresults}
\end{figure*}

It can be seen in FIG. \ref{fig:numericalresults} that for very broad basis functions, except in the case of very few particles, OSE is greatly outperformed both in number of qubits and total tensor weight by the JW mapping. For tight basis functions resulting in local interactions between modes, the performance of OSE is within the same of order of magnitude in terms of total tensor weight as JW. 

Based on the counting of terms, one can determine the required number of qubits for a given lattice size as well as the scaling of the maximum tensor weight term. We have done so for chains of length $N$, and square and cubic grids of side lengths $N$. The results in TABLE \ref{table: scaling} show that at the cost of a constant factor in the number of qubits, the maximum weight of Pauli terms in the transformed Hamiltonian can be bounded by a constant, the graph degree. This is in contrast to being extensive in the size of the lattice as in JW. In higher than one dimension, this will eventually lead to a crossover beyond which OSE mapping will lead to a smaller total tensor weight, however that point appears to be beyond the sizes of systems which we were able to analyze in this study. 

For the Hydrogen systems, the coefficient $L_1$ norms were similar between the two mappings. This is expected given that the term coefficients are changed little by the mappings.  The overall trend in the coefficient norm that we saw was that, as the basis functions were tightened, the norm tended to increase. We did not investigate the effect of the tightened basis sets on the contributions from the kinetic and from the potential terms.

The coefficient $L_1$ norm of the two mappings differed slightly but in most cases only in the second or third decimal place. The notable exception was the smallest system we studied, $H_2$. This is primarily due to the parity constraint of OSE discussed in \ref{sec:empty}. Terms that violate the restriction to the even-particle sector must vanish under OSE mapping.  Since the JW mapping has no such constraint, such terms remain and contribute to the coefficent $L_1$ norm of the Hamiltonian.

\begin{table}
\caption{Scaling of qubit requirement and maximum tensor weight for JW and OSE in simulation of nearest-neighbor lattice models in 1-, 2-, and 3-D. $N$ is length in lattice spacing.}
    \begin{tabular}{c |c c c c}
       \hline
       Geometry & $Q_{JW}$  & $Q_{OSE}$ & Max JW &\phantom{spc} Max OSE\phantom{spc} \\
       \hline
       $N$ &  $2N$ & $2(N-1)$ & $O(1)$ & $O(d)$ \\
       $N\times N$&  $2N^2$ & $4(N^2-N)$ & $O(N)$ & $O(d)$ \\
       $N\times N\times N$&  $2N^3$ &  $6(N^3-N^2)$ & $O(N^2)$ & $O(d)$ \\
       \hline
    \end{tabular}
    \label{table: scaling}
\end{table}

\subsection{AE6 molecules}

\begin{figure}
    \centering
    \includegraphics[width=7cm]{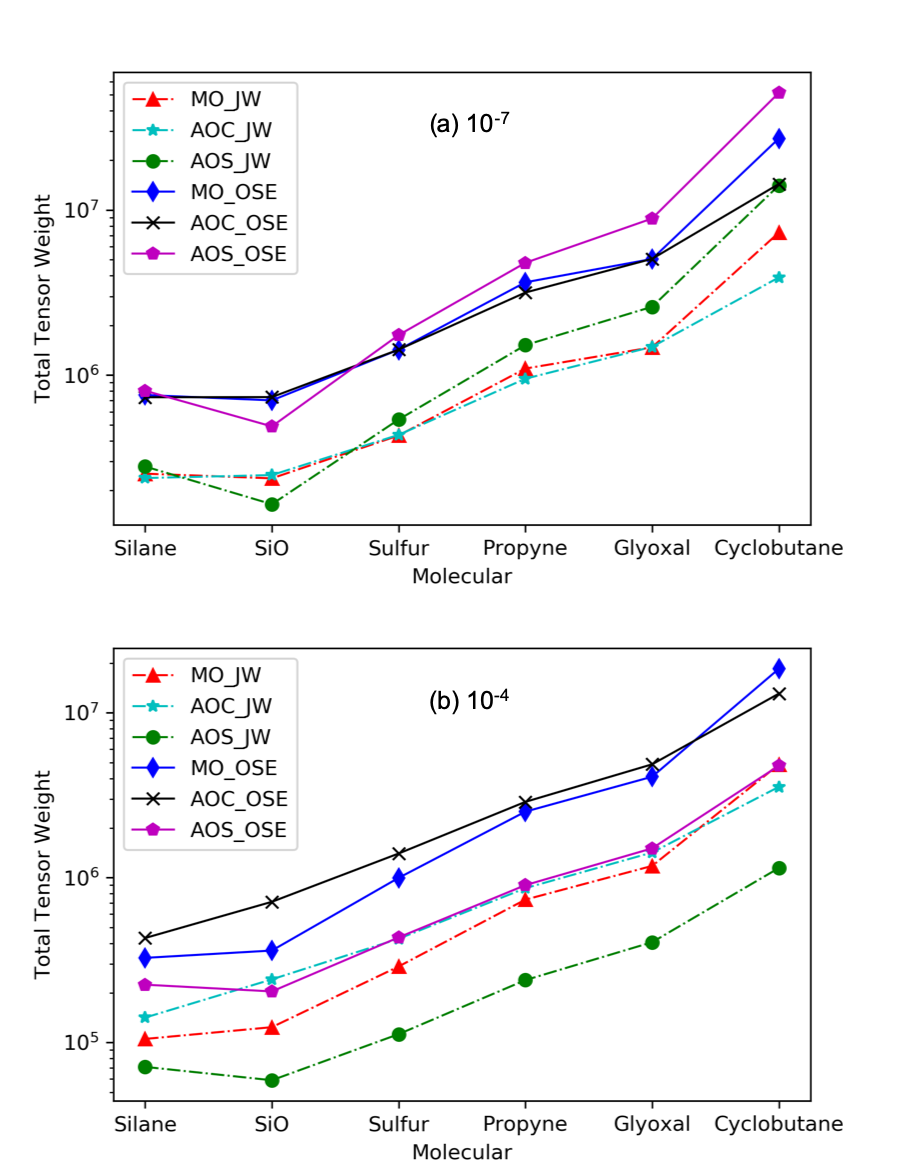}
    \caption{Total tensor weight of all operators in JW and OSE transformed qubit Hamiltonians for the AE6 molecular with different rotation of the basis set (STO-3G). (a): cutoff = $10^{-7}$, (b): cutoff = $10^{-4}$.}
    \label{fig:1e-4}
\end{figure}

In this section, we report results from the study of OSE as applied to the AE6 ensemble. The STO-3G basis set (labeled as Exponent 1 in FIG. \ref{fig:numericalresults} (third column)) were used in the Hartree-Fock calculations to get the one- and two-body integrals. AOC and AOS rotations were also studied in addition to the Hartree-Fock molecular orbitals. FIG. \ref{fig:1e-4} shows the total tensor weight for the different basis set rotations for all six AE6 molecules with different cutoffs. The total tensor weight with cutoff $10^{-7}$ are comparable for AOS, AOC and MO. However, the AOS rotation dramatically reduced the tensor weight for both JW and OSE at cutoff $10^{-4}$. Another major point illustrated by FIG. \ref{fig:1e-4} is that OSE fails to perform better than the JW transformation. With the JW transformation, the number of qubits required is equal to the number of fermion modes present in the system. For example, Silane, SiO, Sulfur, Propyne, Glyoxal, and Cyclobutane need 26, 28, 36, 38, 44, and 56 qubits respectively. For OSE, the number of qubits required depends on the basis set chosen. Upper and lower bounds are compared with JW in TABLE \ref{table:bounds}. Using STO-3G, OSE interaction graphs for each spin sector were fully connected so the qubit number equals the upper bound in TABLE \ref{table:bounds}. 

We considered the effect of localizing basis sets for the AE6 molecules as well. Similarly to the previous section, the investigation of tightened basis functions was not intended to report accurate energies but to investigate the limits of atom-centered orbitals for OSE. With STO-3G basis set, three primitive Gaussian orbitals are fitted to a single Slater-type orbital. Dropping off Gaussian functions with small exponents and choosing larger exponents are two methods to make the orbitals sharper and more localized. Here, we began with the STO-3G orbitals with the most diffuse (exponent 0.169) Gaussian primitive of Hydrogen removed.  Then all remaining Gaussian primitive exponents were scaled uniformly to produce the various orbitals used within our tests.  
We have referred to the basis set used as `Exponent $x$' where $x$ is the multiplicative factor applied to all kept orbital exponents. For example, `Exponent 3' indicates all exponents were tripled. 

As the orbitals become localized (example localized orbital shown in FIG. \ref{fig:Gaussians}), the total energies of the molecules deviate far from the values found when using STO-3G. For example, the total energy for Silane with STO-3G basis is -281.91 Hartrees. However, that value is -194.14 Hartrees with the most localized basis set. FIG. \ref{fig:numericalresults}(i), (g), (k) and (l) show the comparison of metrics which includes total tensor weight, average tensor weight, maximum tensor weight, and $L_1$  between different basis sets with AOS and cutoff $10^{-7}$. As the exponents increase, the total qubits required by OSE transformation for Silane and SiO stayed the same as STO-3G. However, qubits for other molecules were decreased (306 to 302 for Sulfur, 342 to 330 for Propyne, 462 to 432 for Glyoxal and 756 to 708 for Cyclobutane) indicating that some exchange between fermions had been already disappeared. A clearer trend was obtained with cutoff $10^{-4}$. However, even with extremely localized orbitals (Exponent 6), the total qubit number with OSE transformation is still much larger than that of JW. For example, there are 26 orbitals for Silane, therefore 26 qubits are needed with JW transformation. However, for OSE, the number of qubits 136 for OSE even with the most localized basis set we chose. 

As compared to the Hydrogen systems, the trend in the norms was less clear however the tightest basis functions also gave rise to the greatest coefficient $L_1$ norms of the transformed Hamiltonians. The AE6 molecules also saw little difference between JW and OSE. This agrees with the results of the Hydrogen chains, that mapping has a much smaller effect on the norms than on the tensor weight.

\section{Conclusions}

Our conclusions from this paper are twofold. Firstly, the results from the AE6 calculations show that for small molecules using standard basis sets based on atomic orbitals, in particular STO-3G, OSE performs much worse than JW on all metrics of comparison. Indeed, the interaction graph (or pair of graphs - one for each spin sector) on which the transformation is based was in many cases complete. Thus it required many more qubits and longer Pauli terms than are necessary following a JW transformation. We conclude from this, that OSE is not suited to molecular simulations which use standard atomic orbital basis sets. In order to illustrate the dependence on basis function size, we varied the orbital exponent.  While accuracy is lost, the reduction in resource requirements due to the locality of the basis could be observed.

Secondly, the investigation of Hydrogen lattices showed that performance was highly dependent on the width of the single Gaussian orbital used per atom. As with the AE6 results, the broadest orbitals used resulted in a complete or nearly complete interaction graph and thus showed far worse performance for OSE when compared to JW. For tight orbitals, we see performance in the number of qubits and the absolute total tensor weight which is more comparable to JW. The performance was also dependent both on the dimension and size of the lattice.

In 1-D, OSE required two fewer qubits (one for each spin) than JW for all chain lengths and a larger total tensor weight by a factor of up to 1.59 in the largest chain. In 2-D and 3-D, OSE performed worse in all metrics compared to JW for the sizes of systems considered except for maximum tensor weight in 3-D for the largest system considered. The largest 3-D system considered showed a larger total tensor weight by a factor of about $3.61$ while the number of qubits is larger being given by $6(N^3-N^2)$ versus $2N^3$ for JW. Finally, we extrapolated these results to lattices of arbitrary size, $N$ and showed that given a sufficiently large lattice OSE would actually perform better in the total tensor weight. From this, we infer that in the presumptive future regime of large numbers of qubits, a real space tiling of orbitals combined with OSE mapping would be better than other strategies.
We also saw that the choice of mapping has little effect on the coefficient $L_1$ norm of the Hamiltonians therefore tensor weight should be the consideration when choosing a mapping.

In the future, we are planning to extend our inquiry into other fermion to qubit mappings especially the recent generalization of the OSE.  
Additionally, as the quantum device architecture stabilizes, we may soon be able to take the qubit connectivity into account when considering the locality properties of the mappings. We also intend to investigate strategies to mitigate the effects of noise and techniques to overcome device connectivity constraints in conjunction with various mappings \cite{murali2019noise,finigan2018qubit}. Further, we want to consider these mapping in other contexts where fermions play an important role such as high energy physics \cite{Zohar2017,Zohar2018}.  The immediate extension of this work would be to consider basis functions that are not centered on the atom but rather tile space.  In this case, the center to center distances and the width of the orbitals can be traded off to achieve more realistic descriptions of small molecules and metallic Hydrogen.

\begin{acknowledgments}
The authors would like to thank the U.S. Department of Energy for supporting this work as part of the Quantum Computation for Quantum Chemistry (QCQC) Collaboration (Grant No. DE-SC0019374).  JDW is also supported by the NSF under grant PHYS-1820747 and by the U.S. Department of Energy, Office
of Science, Office of Advanced Scientific Computing Research, under the Quantum
Computing Application Teams program (Award 1979657).
\end{acknowledgments}

\end{document}


\section{Appendix 2: AE6 molecules numerical results with cut-off $10^{-4}$}

See Figure 6.

\begin{figure*}
    \centering
    \includegraphics[width=17cm]{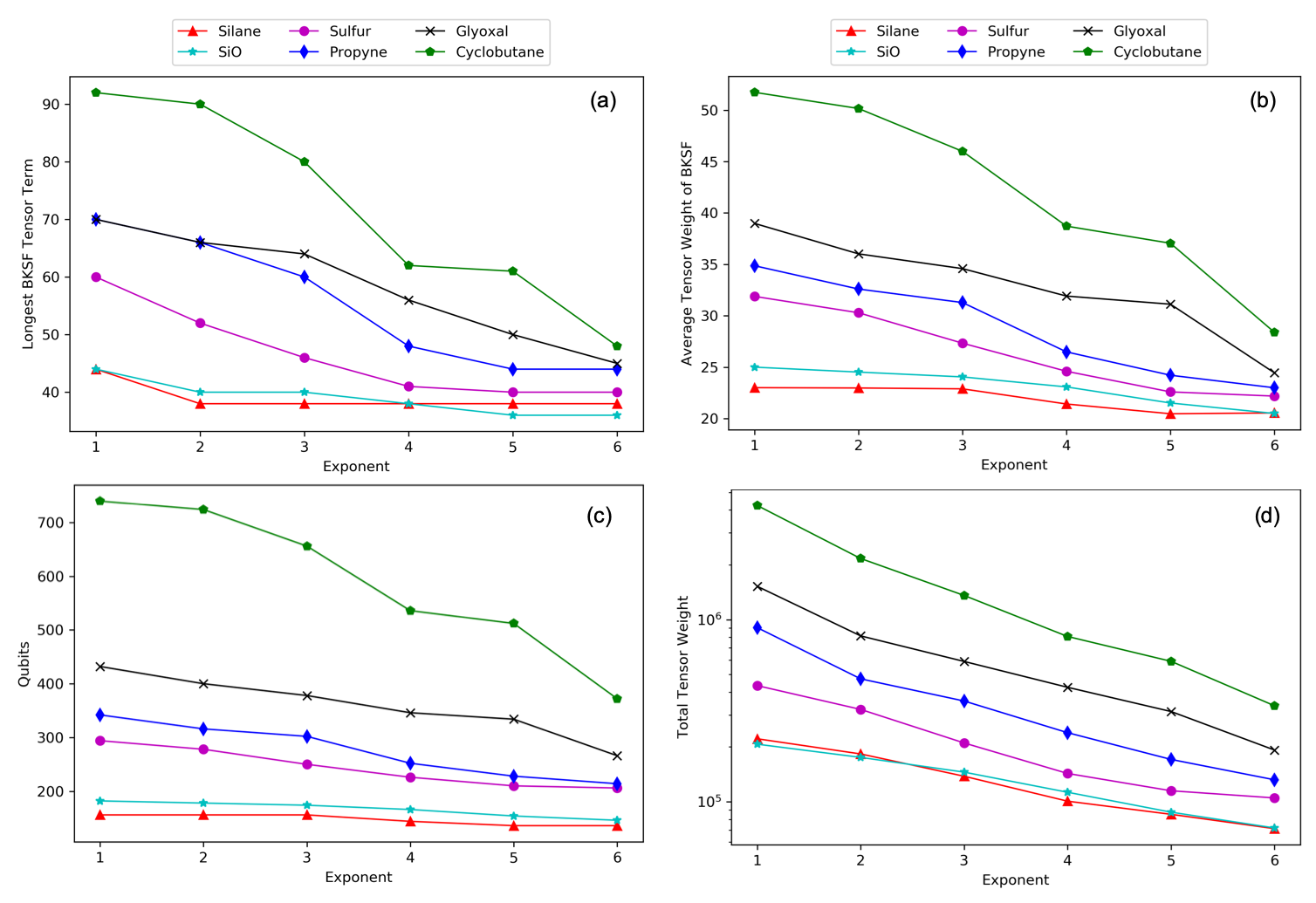}
    \caption{Numerical results (Total Tensor Weight, Average Tensor Weight, Qubits and Longest Tensor Term in OSE transformed qubit Hamiltonians with different customized basis sets) for AE6 molecules with AOS rotation and cutoff $10^{-4}$}
    \label{fig:total tensor weight}
\end{figure*}


\section{Appendix 3: H Lattice Data}

The following is the qubit requirements and total tensor weights of the electronic Hamiltonian following a JW and BKSF transformation. The spacing between each nearest neighbor Hydrogen atom is $1$ Angstrom. A cutoff of $10^{-7}$ was chosen so any one- or twobody integral which fell below that value was neglected. Integrals were done in PSi4 and mapping to qubit operators was done in OpenFermion.

\subsection{1D}
\begin{center}
\begin{tabular}{c|cccc}
Basis=8.75 \\
\hline 
Size & JW Qbts & BKSF Qbts & JW TWt & BKSF TWt\\
\hline
2 & 4 & 2 & 24 & 4 \\
4 & 8 & 6 & 88 & 92 \\
6 & 12 & 10 & 184 & 256 \\
8 & 16 & 14 & 312 & 472 \\
10 & 20 & 18 & 472 & 752 \\
\hline
\end{tabular}

\begin{tabular}{c|cccc}
Basis=7.00 \\
\hline 
Size & JW Qbts & BKSF Qbts & JW TWt & BKSF TWt\\
\hline
2 & 4 & 2 & 48 & 4 \\
4 & 8 & 6 & 160 & 172 \\
6 & 12 & 10 & 304 & 400 \\
8 & 16 & 14 & 480 & 680 \\
10 & 20 & 18 & 688 & 1024 \\
\hline
\end{tabular}

\begin{tabular}{c|cccc}
Basis=5.00 \\
\hline 
Size & JW Qbts & BKSF Qbts & JW TWt & BKSF TWt\\
\hline
2 & 4 & 2 & 48 & 4 \\
4 & 8 & 10 & 328 & 448 \\
6 & 12 & 18 & 784 & 1584 \\
8 & 16 & 26 & 1272 & 2880 \\
10 & 20 & 34 & 1792 & 4304 \\
\hline
\end{tabular}

\begin{tabular}{c|cccc}
Basis=3.00 \\
\hline 
Size & JW Qbts & BKSF Qbts & JW TWt & BKSF TWt\\
\hline
2 & 4 & 2 & 64 & 4 \\
4 & 8 & 12 & 632 & 928 \\
6 & 12 & 24 & 1504 & 3552 \\
8 & 16 & 36 & 2600 & 7144 \\
10 & 20 & 48 & 3920 & 11696 \\
\hline
\end{tabular}

\begin{tabular}{c|cccc}
Basis=1.00 \\
\hline 
Size & JW Qbts & BKSF Qbts & JW TWt & BKSF TWt\\
\hline
2 & 4 & 2 & 68 & 4 \\
4 & 8 & 12 & 1440 & 1768 \\
6 & 12 & 30 & 7504 & 15136 \\
8 & 16 & 56 & 19368 & 51376 \\
10 & 20 & 90 & 33528 & 113340 \\
\hline
\end{tabular}
\end{center}

\subsection{2D}
\begin{center}
\begin{tabular}{c|cccc}
Basis=8.75 \\
\hline 
Size & JW Qbts & BKSF Qbts & JW TWt & BKSF TWt\\
\hline 
4 & 8 & 8 & 104 & 128 \\
16 & 32 & 48 & 1360 & 3280 \\
\hline
\end{tabular}

\begin{tabular}{c|cccc}
Basis=7.00 \\
\hline 
Size & JW Qbts & BKSF Qbts & JW TWt & BKSF TWt\\
\hline
4 & 8 & 8 & 216 & 256 \\
16 & 32 & 48 & 2224 & 4496 \\
\hline
\end{tabular}

\begin{tabular}{c|cccc}
Basis=5.00 \\
\hline 
Size & JW Qbts & BKSF Qbts & JW TWt & BKSF TWt\\
\hline
4 & 8 & 8 & 424 & 384 \\
16 & 32 & 80 & 11008 & 26862 \\
\hline
\end{tabular}

\begin{tabular}{c|cccc}
Basis=3.00 \\
\hline 
Size & JW Qbts & BKSF Qbts & JW TWt & BKSF TWt\\
\hline
4 & 8 & 12 & 928 & 1208 \\
16 & 32 & 180 & 25464 & 116912 \\
\hline
\end{tabular}

\begin{tabular}{c|cccc}
Basis=1.00 \\
\hline 
Size & JW Qbts & BKSF Qbts & JW TWt & BKSF TWt\\
\hline
4 & 8 & 12 & 1376 & 1704 \\

16 & 32 & 240 & 456736 & 1687072 \\
\hline
\end{tabular}
\end{center}

\subsection{3D}
\begin{center}
\begin{tabular}{c|cccc}
Basis=8.75 \\
\hline 
Size & JW Qbts & BKSF Qbts & JW TWt & BKSF TWt\\
\hline
8 & 16 & 24 & 416 & 864 \\
64 & 128 & 288 & 20992 & 75840 \\
\hline
\end{tabular}

\begin{tabular}{c|cccc}
Basis=7.00 \\
\hline 
Size & JW Qbts & BKSF Qbts & JW TWt & BKSF TWt\\
\hline
8 & 16 & 24 & 832 & 1440 \\
64 & 128 & 288 & 31360 & 86592 \\
\hline
\end{tabular}

\begin{tabular}{c|cccc}
Basis=5.00 \\
\hline 
Size & JW Qbts & BKSF Qbts & JW TWt & BKSF TWt\\
\hline
8 & 16 & 24 & 3200 & 4704 \\
64 & 128 & 480 & 374368 & 731136 \\
\hline
\end{tabular}

\begin{tabular}{c|cccc}
Basis=3.00 \\
\hline 
Size & JW Qbts & BKSF Qbts & JW TWt & BKSF TWt\\
\hline
8 & 16 & 24 & 6896 & 16896 \\
64 & 128 & 1584 & 938464 & 4989312 \\
\hline
\end{tabular}

\begin{tabular}{c|cccc}
Basis=1.00 \\
\hline 
Size & JW Qbts & BKSF Qbts & JW TWt & BKSF TWt\\
\hline
8 & 16 & 56 & 35072 & 84832 \\
64 & 128 & 3976 & 37685904 & 235369360 \\
\hline
\end{tabular}

\end{center}
\vspace{10mm}